\documentclass[a4paper]{jpconf}
\usepackage{graphicx}
\begin{document}
\title{Probing the correlations in composite signals}

\author{S. Sargsyan}

\address{Center for Cosmology and Astrophysics, Alikhanian National Laboratory and Yerevan State University, Yerevan, Armenia}

\ead{seda@yerphi.am}

\begin{abstract}
The technique of degree of randomness is used to model the correlations in sequences containing various subsignals and noise. Kolmogorov stochasticity parameter enables to quantify the randomness in number sequences and hence appears as an efficient tool to distinguish the signals. Numerical experiments for a broad class of composite signals of regular and random properties enable to obtain the qualitative and quantitative criteria for the behavior of the descriptor depending on the input parameters typical to  astrophysical signals.     
\end{abstract}

\section{The method}

The correlations functions and the power spectra are common and efficient tools for the study of correlations in the signals. Astrophysical signals typically are superposition of various subsignals, regular and random, by features comparable to each other and of weaker ones, i.e. perturbations or the noise. The procedure of analysing of the needed signal or signals, their separation from the noise is a common problem while dealing with observations and measurements. 

The Kolmogorov stochasticity parameter technique enables to quantify the randomness of sequences of number theory or dynamical systems \cite{Kolm,Arnold,Arnold_UMN,Arnold_MMS,Arnold_FA}. 

The technique of the degree of randomness has been applied to the Cosmic Microwave Background (CMB) temperature sky maps and to the X-ray flux data of the clusters of galaxies. The former data were those obtained by the Wilkinson Microwave Anisotropy Probe (WMAP) during 7-year observations \cite{K,J}, while the X-ray data were obtained by XMM-Newton satellite providing a particularly accurate and complete sky survey (see \cite{XMM,XMM1}).  In the case of CMB, the Kolmogorov function enabled to separate signals of different origin, e.g. the Galactic and non-Galactic ones, and to detect point sources in the CMB maps \cite{G2009} (see Fig.1). Concerning the X-ray clusters, it was shown that their X-ray images do possess correlation in the pixelized flux data peculiar  to the gravitational potential of the galaxy clusters \cite{GD2011}. This technique resembles the methods of the dynamical systems applied to nonlinear problems (e.g. \cite{GP}).

A crucial step in these studies is the modeling and analysis of generated systems, which enables to reveal the behavior of the stochasticity parameter in the case of a given signal and then to consider the application of this technique for real signals \cite{EPL2011}. Below we represent the results of numerical experiments for a broad class of signals.  

Kolmogorov stochasticity parameter is introduced for a sequence $\{X_1,X_2,\dots,X_n\}$ of  real random variable $X$ sorted in growing order  $X_1\le X_2\le\dots\le X_n$. Then the theoretical distribution function is \cite{Kolm,Arnold}
\begin{equation}
F(X) = n \cdot(probability\, of\, the\, event\, x \leq X). 
\end{equation}
the stochasticity parameter is defined as 
\begin{equation}\label{KSP}
\lambda_n=\sqrt{n}\ \sup_x|F_n(x)-F(x)|\,
\end{equation}
where the empirical distribution function is
$$F_n(X)= (number\, of\, the\, elements\, x_i\, which\, are\, less\, than\, X),$$
and
\begin{equation}
F_n(X)= \left\{
\begin{array}{rl}
	0, & X < x_1 \\
	k / n, & x_k \leq X < x_{k+1} \\
	1, & x_n \leq X.\\
\end{array}
\right.
\label{eq:empiricdistribution}
\end{equation}

Then for the limit 
\begin{equation}
\lim_{n\to\infty}P\{\lambda_n\le\lambda\}=\Phi(\lambda)\ ,
\end{equation}
where
\begin{equation}
\Phi(\lambda)=\sum_{k=-\infty}^{+\infty}\ (-1)^k\ e^{-2k^2\lambda^2},\, \Phi(0)=0,\,   \lambda>0\ ,\label{Phi}
\end{equation}
exists at uniform convergence and independent on $F$.

\begin{figure}[h]
\includegraphics[width=20pc]{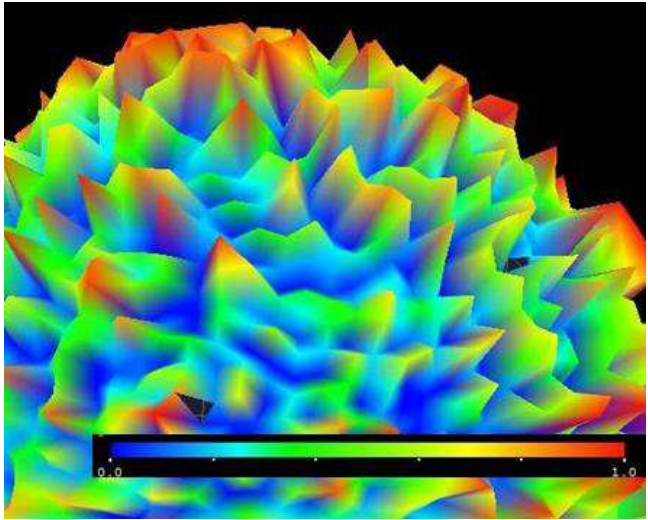}\hspace{2pc}%
\caption{\label{fig:phi1}The Kolmogorov's function $\Phi$ for the portion of the 7-year temperature CMB map obtained by WMAP.}
\end{figure}

\section{Random-regular sequences}

We consider a broad class of sequences, i.e. those composed of random $x_n$ and regular $y_n = \frac{an\pmod b}{b}$ ($a,b$ are prime numbers) sub-sequences within $(0,1)$ 
\begin{equation}
z_n = \alpha x_n + (1-\alpha) y_n.
\end{equation}
 
The parameter $\alpha$ varies within [0,1] defining random sequences at $1$ and regular ones at $0$, so that by mutually fixing the pair $a,b$ we get new regular sequences. 

For $z_n$ we have
\begin{equation}
F(X)= \left\{
\begin{array}{rl}
	0, & X \leq 0 \\
	\frac{X^2}{2 \alpha (1-\alpha)}, & 0 < X \leq \alpha\\
	\frac{2 \alpha X - \alpha^2}{2 \alpha (1-\alpha)}, & \alpha < X \leq 1-\alpha\\
	1-\frac{(1-X)^2}{2 \alpha (1-\alpha)}, & 1-\alpha < X \leq 1\\
	1, & X > 1.\\
\end{array}
\right.
\label{eq:d}
\end{equation}

Figure 2 shows the results of the numerical experiments for 100 sequences, each sequence containing 10000 elements.
Each sequence is divided into 50 subsequences, i.e. $m$ runs through values $1, ..., 50$, and for each of them
the parameter $\Phi(\lambda_n)_m$ is calculated and then the empirical distribution function $G(\Phi)_m$ of these 
numbers is obtained. When the original sequences are random, this distribution have to be uniform according to Kolmogorov's theorem. To test that, $\chi^2$ for the functions $G(\Phi)_m$ and $G_0(\Phi)=\Phi$ have been calculated, i.e. one 
parameter $\chi^2$ is calculated for each of the $100\times101$ sequences. For $100$ $\chi^2$ values per each value of $\alpha$, we obtained 
the mean and error values for $\chi^2$, i.e. for each pair $a,b$ we have a plot of the dependence of $\chi^2$ on $\alpha$.  

\begin{figure}[h]
\includegraphics[width=25pc]{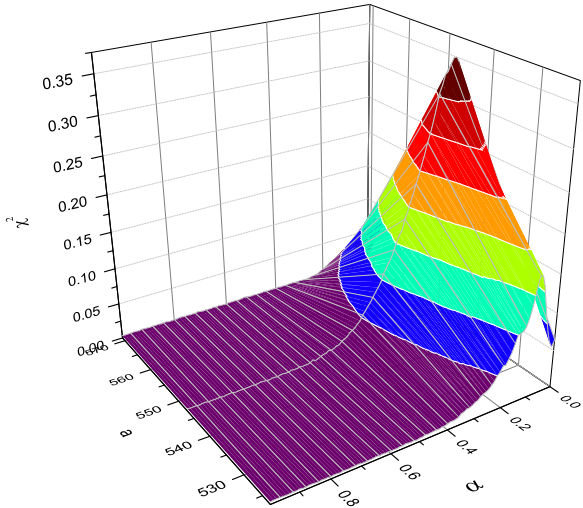}\hspace{2pc}%
\caption{\label{fig:chi_sq}The 3D $\chi^2$ for the Kolmogorov's function for the sequence $z_n$ vs $\alpha$ and the parameter $a$.}
\end{figure}

\section{Parameters of the regular sequences} 

At certain values of the parameter $a$ for different values of $b$ the dependences in Fig.2 are monotonic, while for others they do have maxima. To study this effect, we introduce a parameter $\Delta$ which is the difference of two values in those plots: maximal value of $\chi^2$ and minimal value in the range $\alpha \in (0, \alpha_{max})$, 
if $\alpha_{max}$ is the position of the maximal value. Obviously, $\Delta$ is zero when the dependence is monotonic and no extrema do exist. Then we calculate $\Delta$ for fixed $b$ and for each value of primary $a={2, ..., b}$ . 

\begin{figure}[ht]
\includegraphics[width=20pc]{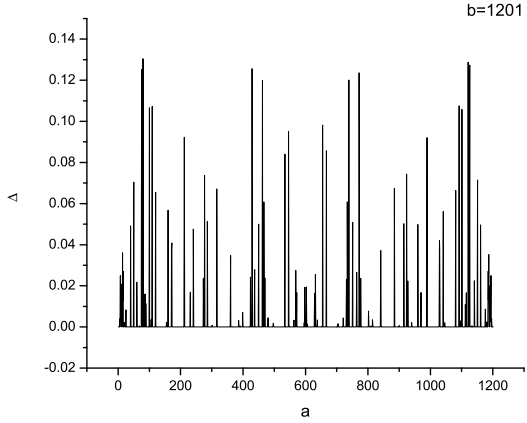}\hspace{2pc}%
\caption{\label{fig:deltas}The dependence of $\Delta$ vs the parameter $a$.}
\end{figure}
The remarkable feature of the results is the strict mirror symmetry in Fig.\ref{fig:deltas} in the dependence of $\Delta$ vs $a$, although no periodicity is found by Fourier analysis.

The mirror symmetric plots can be hence subjects of particular study, e.g. in two versions: first, of the distribution of $\Delta$ and, second, the spacing between non-zero $\Delta$ and their distribution.
The null values of $\Delta$ are skipped and also - since due to the mirror symmetry each $\Delta$ has its pair - only one of each pair is taken into account. 

\begin{figure}[ht]
\includegraphics[width=20pc]{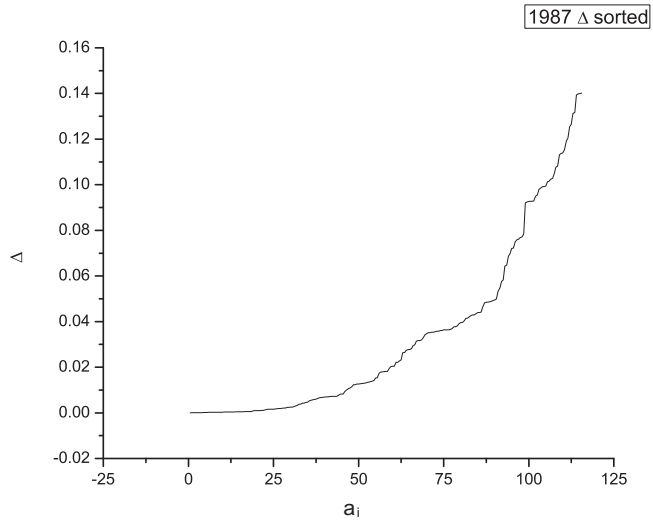}\hspace{2pc}%
\caption{\label{fig:y1}Sorted amplitudes of $\Delta$s from Fig.\ref{fig:deltas}.}
\end{figure}

The results we give in Fig.\ref{fig:y1} where the number of non-zero $\Delta$s from Fig.\ref{fig:deltas} is given in growing order. The number of non-zero $\Delta$s appear to be proportional to $b$.

\section{Sum of fluctuations: large N limit}

The next problem we consider is the properties of the signal being a sum of random and regular fluctuations,   
each of sequences of 10000 elements and of the same standard deviation.

\begin{figure}[ht]
\includegraphics[width=14pc]{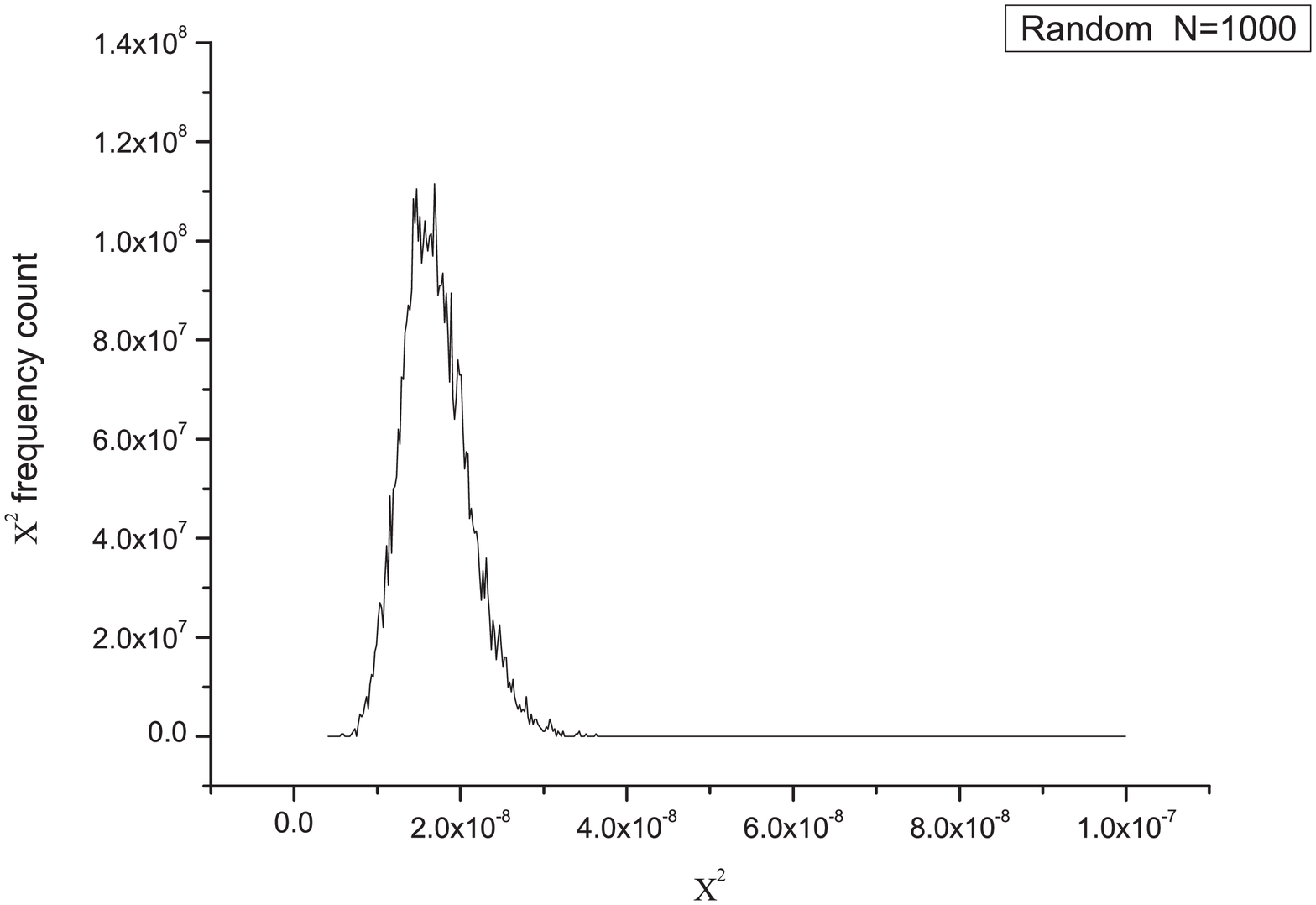}\hspace{2pc}%
\includegraphics[width=14pc]{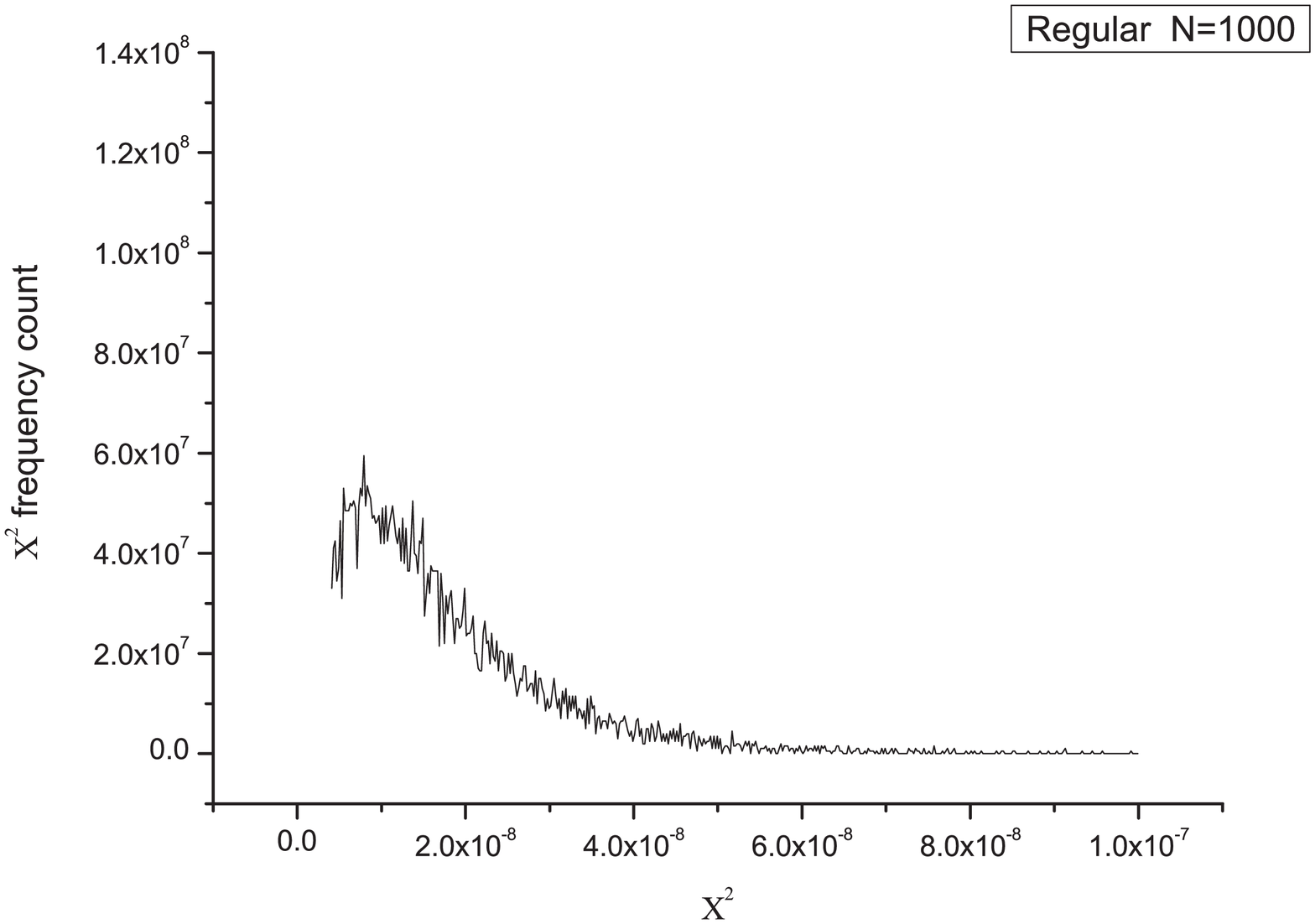}\hspace{2pc}%
\caption{\label{fig:chi_dist}$\chi^2$ frequency for random and regular sequences vs 
the Gaussian distribution.}
\end{figure}

\begin{figure}[ht]
\includegraphics[width=14pc]{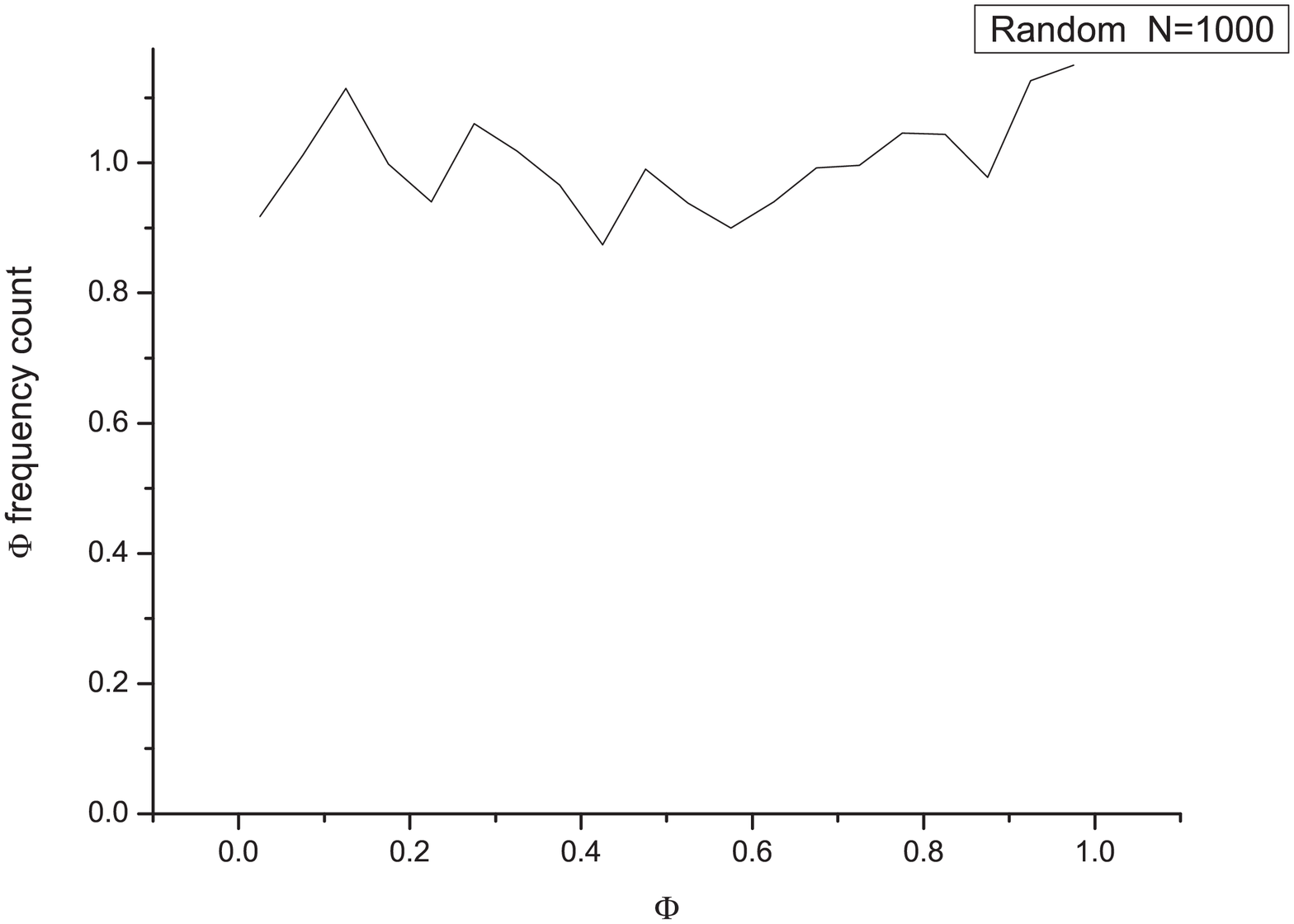}\hspace{2pc}%
\includegraphics[width=14pc]{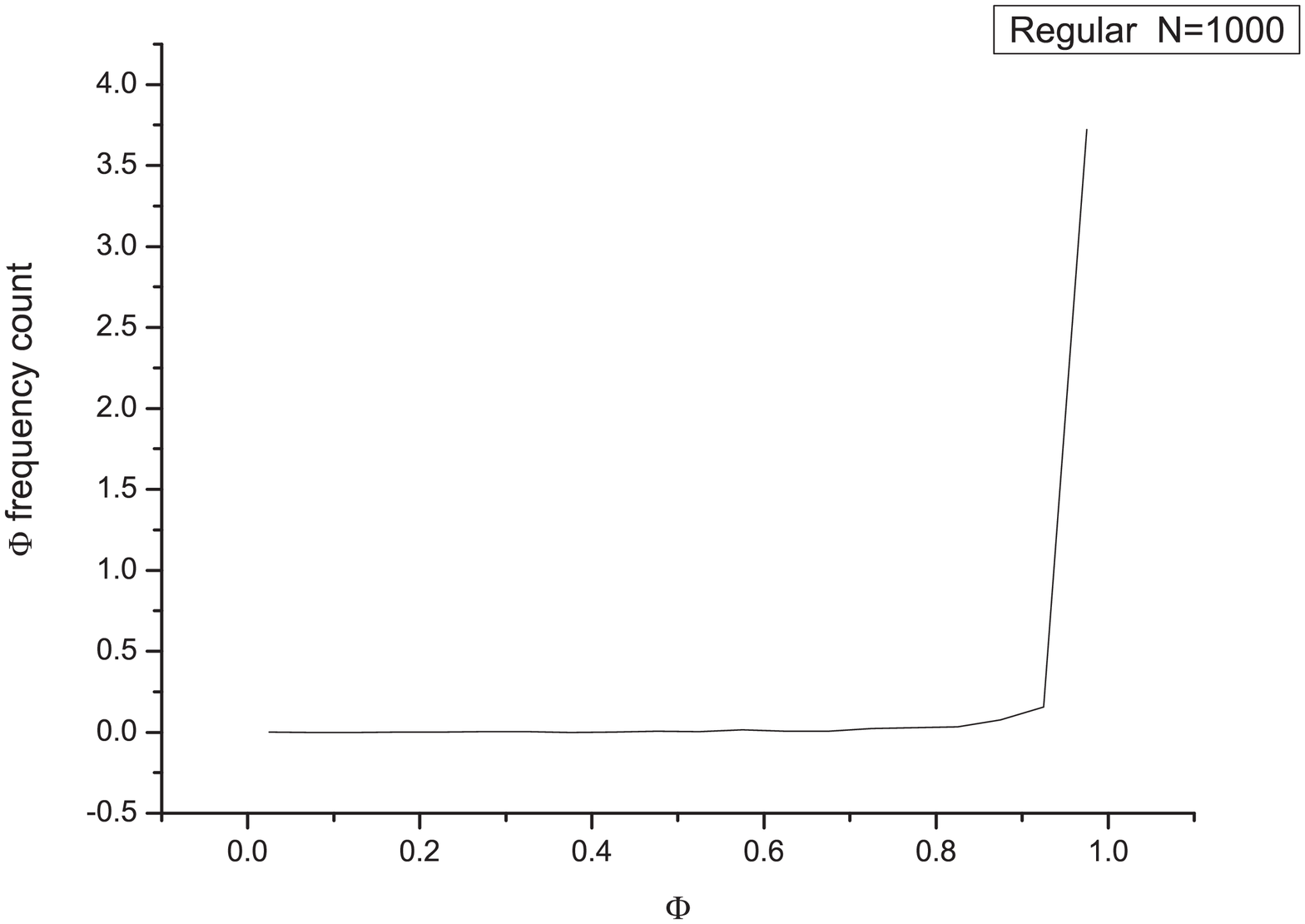}\hspace{2pc}%
\caption{\label{fig:phi}Kolmogorov function $\Phi$ for the random and regular sequences in Fig.\ref{fig:chi_dist}.}
\end{figure}

The regular sequences have been chosen as
\begin{equation}
a_i=\frac{1}{\sqrt N} \sum_{k=1}^N Compact(x_i^k,-1,1),
\end{equation}
where $$x_i^k=i/k,$$
$x^k$ is compactified arithmetical sequence within the 
interval $(-1,1)$, with step $1/k$. 

The random sequences are given by
$$
b_i=\frac{1}{\sqrt N} \sum_{k=1}^N Random(-1,1).
$$ 
At large number of sequences each new sequence $y_n$ is taken as the continuation of the former arithmetical progression.

Here 
$$Compact(x,p,q)=q+x mod(p-q)$$ 
indicate multiples of $(p-q)$ from $x$ having the value within the range $(p,q), p<q$. 

The results for random  and regular sequences, 10000 each, are given in Fig.\ref{fig:chi_dist} for $\chi^2$, 
when the number of the fluctuations vary from $N=1000$ to $100000$.
The $\chi^2$ shows that for both, random and regular sequences, we deal with a Gaussian
limiting distribution, in accordance with the Central Limit theorem which states that for large enough values of $N$, both sequences $a_i$ and 
$b_i$ tend to Gaussian sequences with the same $\sigma$ and $\mu$ independent on $N$. So, although differences are seen in the 
Gaussians, namely, the standard deviations are larger for the regular case, the $\chi^2$ are similar. 

For the Kolmogorov function $\Phi$ the situation is rather different. When the Gaussians do appear both for random and regular sequences (as expected), the behaviors of $\Phi$ is different and enables to separate them, as shown in Fig.\ref{fig:phi}. Namely, it is close to a homogeneous function for random sequences and $\Phi=1$ for regular ones. Kolmogorov's function therefore enables to distinguish the superposition of random and regular sequences, even though both are tending to Gaussians.  
  
Finally, we have probed the dependence of the results on the length of the sequences: the dependence on the number of the fluctuations within $1000-100000$ is rather weak, $\chi^2$ varying around $10^{-8}-10^{-9}$. This confirms the universality of the obtained behavior of $\Phi$ for both random and regular fluctuations.

\section{Results}

The performed analysis revealed the behavior of the Kolmogorov distribution vs the properties of the generated signals. To describe datasets which contain both regular and stochastic components, we considered sequences scaled by a single parameter $\alpha$, indicating the ratio of those components. 

Quantitative and qualitative criteria have been obtained for the Kolmogorov distribution at numerical experiments for broad class of random and regular sequences depending on $\alpha$ parameter. 

a) The existence of the {\it critical value} for $\alpha$ has been shown, when the monotonic decay 
of the frequency count of the Kolmogorov distribution is transformed to a function with an extremum. 

b) The dependence of scalings and spacings vs that parameter 
shows mirror properties both in the amplitude and distribution of the frequency counts of the function $\Phi$. 

c) The behavior of the randomness of a signal composed of $N$ subsignals at large $N$ limit has been studied, where the Kolmogorov function acts as an informative descriptor. Particularly, the descriptor at large $N$ enables to distinguish the initial set of the fluctuations, even when the superposition both of random and regular subsignals is not informative since tends to a Gaussian in accordance to the Central Limit theorem. 

The studied properties are typical, for example, for astrophysical datasets, when the sought signals are superposed with regular and random fluctuations of various origin, and hence the behaviors revealed at the numerical experiments due to the universality of the technique will enable its informative application to real data.

\smallskip 

\medskip
\begin{thebibliography}{12}

\bibitem{Kolm}
Kolmogorov A.N. 1933 {\it G.Ist.Ital.Attuari,} {\bf 4} 83 

\bibitem{Arnold}
Arnold V.I. 2008 {\it Nonlinearity} {\bf 21} T109 

\bibitem{Arnold_UMN}
Arnold V.I. 2008 {\it Uspekhi Mat. Nauk} {\bf 63} 5 

\bibitem{Arnold_MMS}
Arnold V.I. 2009 {\it Trans. Mosc. Math. Soc.} {\bf 70} 31 

\bibitem{Arnold_FA}
Arnold V.I. 2009 {\it Funct. Anal. Other Math.} {\bf 2} 139 

\bibitem{K}
Komatsu E., Dunkley J. {\it et al.} 2009 {\it ApJS} {\bf 180} 330 

\bibitem{J}
Jarosik N., Bennett C.L. {\it et al.} 2011 {\it ApJS} {\bf 192} 14 

\bibitem{XMM}
Viana P.T.P., da Silva A.  {\it et al.} 2011 {\it arXiv:1109.1828}

\bibitem{XMM1}
Suhada R., Song J., {\it et al.} 2011 {\it arXiv:1111.0141} 

\bibitem{G2009}
Gurzadyan V.G., Allahverdyan A.E. {\it et al.} 2009 {\it Astron. \& Astrophys.} {\bf 497} 343

\bibitem{GD2011}
Gurzadyan V.G., Durret F.  {\it et al.} 2011 {\it Europhys.Lett.} {\bf 95} 69001

\bibitem{GP} Gurzadyan V.G., Pfenniger D., (Eds.) 1994 {\it Ergodic Concepts in Stellar Dynamics}, Springer-Verlag.

\bibitem{EPL2011}
Gurzadyan V.G., Ghahramanyan T., Sargsyan S. 2011 {\it Europhys.Lett.} {\bf 95} 19001

\end{thebibliography}
\end{document}